\documentclass[runningheads,orivec]{llncs}
\usepackage[T1]{fontenc}
\usepackage{amsmath}
\usepackage[utf8]{inputenc}
\usepackage{xcolor}
\usepackage{url}
\usepackage{listings}
\usepackage{hyperref}
\hypersetup{
    colorlinks,
    linkcolor={red!50!black},
    citecolor={blue!50!black},
    urlcolor={blue!80!black}
}
\usepackage{cite}  
\usepackage{array} 
\usepackage{comment}
\newcommand{\sys}[1]{{\sc #1}}
\newcommand{\Cpp}{\sys{C}\texttt{++}}
\newcommand{\spacetune}[1]{#1}

\title{Software Portability for Computer Algebra}
\author{
  Arthur C. Norman\inst 1 
  \and 
  Stephen M. Watt\inst 2
}
\institute{
  Trinity College,
  Cambridge CB2 1TQ, UK\\
  \email{acn1@cam.ac.uk}\\
  ~
  \and
  Cheriton School of Computer Science,
  University of Waterloo, N2L 3G1 Canada\\
  \url{https://cs.uwaterloo.ca/~smwatt} \\
  \email{smwatt@uwaterloo.ca}
}
\date{\today}
\begin{document}
\maketitle
\begin{abstract}
  We have been involved in the creation of multiple software systems for computer algebra, including \sys{Reduce}, \sys{Maple}, \sys{Axiom} and \sys{Aldor} as well as a number of smaller specialised programs.   
  We relate observations on how the meaning of software portability has changed over time and how it continues to evolve.  
  We describe how the systems with which we have first-hand experience have achieved portability,
  how the central issues have changed over time
  and the challenges that remain.
\end{abstract}
\section{Introduction}
\label{sec:introduction}
This article lays out the issues we have observed in the creation of computer algebra systems, and the open issues we see remaining.
Software systems for computer algebra have been available since the 1960s on a wide variety of computing platforms.   
The underlying architectures of these platforms varied in major respects and numerous challenges had to be overcome to allow a given computer algebra system to operate in different environments.
Computer algebra programs have characteristics that place particular demands on the supporting software: dynamic memory allocation was required to handle a great many small, mostly read-only, data objects with varying lifetimes and the efficiency of certain key data types (such as arbitrary precision integers or integers mod a prime) was absolutely critical to overall system performance.
Another feature of computer algebra systems is their embodiment of multiple sophisticated algorithms, each involving numerous special cases (for efficiency) and implemented sitting atop a library of equally elaborate sub-algorithms.   These algorithms cannot be readily moved from one system to another, so systems tend to be long-lived, spanning multiple generations of computing environments.
For these reasons, the software portability issues for computer algebra have a different emphasis than  is found elsewhere.  

This article is organised as follows:
The meaning of portability is explored in 
Section~\ref{sec:meaning}.
The principal questions arising in early portability of computer algebra systems are 
discussed in
Section~\ref{sec:earlyportability}.
Experience in the portability of \sys{Lisp} systems
over time is related in 
Section~\ref{sec:lispevolve}.
Next,
Sections~\ref{sec:case-reduce}
and~\ref{sec:case-aldor}
present case studies with two approaches to 
computer algebra system portability.
The role of software standards is discussed in
Section~\ref{sec:standards}
and
the idea of portability of mathematical formulations
is considered in
Section~\ref{sec:portable-math}.
Final thoughts and conclusions are given in
Section~\ref{sec:conclusions}.

\section{The Meaning of Portability}
\label{sec:meaning}
Portability of the code that makes up a system at first seems like a simple concept even if in practise achieving it can be challenging. However in the discussions we had leading to this paper we realised gradually more sharply how many facets it has and how the relative importance of them have changed over our working lives. So here we will cite examples to illustrate some of these, and we will then provide more in-depth commentaries on how they have impacted particular computer algebra software and the steps that were taken or might be taken to respond to them.

In the early days of computer algebra (notably much of the 1960s) serious scale algebraic computation strained (at the very least!) machine memory capacity, and porting only made sense if the new target hardware was of a style installed at multiple or particularly important sites. A simplistic characterisation was that only the larger machines from IBM, DEC and Cray mattered. But the designs from those three major vendors had little in common either in hardware architecture or standard software. Programming language standardisation emerged through the 1960s with \sys{Fortran}\cite{fortran66} leading the way and \sys{Cobol} following (though probably little serious computer algebra has used \sys{Cobol}), but even beyond the release of the standard it would both be necessary to make careful checks about the compliance level of a local compiler and to be quite thoughtful about the extent to which tempting local extensions would make code run faster or better. The other viable path was to use \sys{Lisp} where although there was good agreement about the specification of its core, versions at different sites or on different computers were enthusiastically enhancing and adjusting their version for local needs. The section we have later on using \sys{Reduce} as a case-study may give more insight into those past and 
dark  
times.

A number of years later Unix had become a dominant operating system for use in the sort of research environment where computer algebra systems were being developed. \sys{C} was ubiquitous and \sys{Lisp} had gained a dominant position for symbolic computation including but going well beyond algebra. However there had been an explosion in the number of vendors of workgroup-scale computers and neither Unix nor \sys{C} nor \sys{Lisp} had internationally accepted standards that could be relied on. Anybody who lived through those decades is likely to recall that making \sys{C} code portable involved scattering horrendous numbers of ``\verb+#ifdef+'' directives through the code. Some would be to allow for operating environment issues, others for word-length, byte order, floating point representation and other hardware matters, or because one idiom was much faster on platform A than B while the situation was reversed for another code fragment that was seen as performance critical. Finally, it was not uncommon to need to use conditional compilation to sidestep bugs in the compiler or some of the libraries that were to be used. \sys{Maple}~\cite{Maple} is an example of a system that emerged against this background. Its history illustrates the challenge of the cost of the computers used elsewhere to develop computer algebra, leading the \sys{Maple} designers to be keen to exploit smaller and more affordable options. It also shows how they based what they were doing on \sys{C} (with very initial work using \sys{B}, its predecessor language). And despite the fact that \sys{C} had an \textit{ad hoc} rather than a formal standard and no doubt with a serious amount of conditional compilation they made their system so it could be transplanted to a range of Unix platforms: portability was in their minds from the start -- indeed, it was in the title of the first \sys{Maple} paper~\cite{Maple83}.

Towards the end of the 1980s a standard for \sys{C} emerged and \sys{Posix}\cite{POSIX} at least explained what a Unix system ought to support. Reworking code to be in \sys{Ansi C} and so that it relied on \sys{Posix} promised to greatly ease the task of migrating code between machines from different vendors --- and this just in time to see the start of a need to cope with the emerging personal computers. These had become powerful enough that an individual scientist could often achieve as much using one that was their own, and could be left running overnight if necessary, as they could on a more powerful departmental or institutional machine. In our discussion of \sys{Aldor}\cite{Aldor-CAH} and the \sys{Axiom}~\cite{Axiom} system it was associated with, we note that at the time NAG Ltd took over responsibility for marketing \sys{Axiom}, it could run on personal-scale hardware at least in cases where the machine was generously (for the time) endowed with memory. So as variants on Linux really started to converge making life easier for developers and maintenance teams, the joys of Windows and the Macintosh emerged and the feeling that users now needed comfortable graphical interfaces (where portable support was even more ``exciting'') arose.

The last remark above highlights a separate issue that can be held to fall under the banner of portability. Early systems of necessity used text-based interfaces\footnote{Apart from some really good early work at MIT and Xerox PARC where special hardware was available to the few.}. When those systems moved into a modern world it was necessary to enhance them with GUIs. While from some perspective one can views this as an extension rather than as porting, our view that it falls into the same challenge and the workload will tend to fall on the same sort of people who would migrate the code to a new
computer. It is porting across the boundaries of time and user expectation rather than across boundaries of operating system dependencies. Other rather similar challenges arose with a need for web-based interfaces and it is plausible that two pressing ones today are to find a response to generative AI
and to provide full exploitation of the concurrency opportunities of cloud-based servers. Each of these can call for re-work deep into the core of a project.

Just when you had hoped that portability challenges were receding into the past, with the transition from 32-bit word length to 64-bits having been one of the more notable (that caused some entertainment for those \sys{Reduce} developers who were concerned with the lowest level of the software stack), one finds two disruptions. The first is when major commonly used brands of computer change the style of CPU they use (\textit{e.g.} Apple changing from PowerPC to Intel to ARM).
Supporting the new situation will of course be merely a variant on the traditional ``port to a different system'' activity, so those who had been active in the 1980s will cope well. However experience shows that support libraries and the like for mildly fringe work such as that on computer algebra systems can take a significant time to stabilise. Also re-working build scripts and the schemes for packing software for distribution such that recipients who have both old and new architectures both have a seamless experience takes effort and sometimes really detailed work. Of course the computer vendors in presenting their transition tend to promise that the adjustments will be really easy to make: experience qualifies this stance --- it is liable to theoretically easy but getting everything sorted out can take an astonishing amount of time.

A more serious ``porting'' challenge can arise when operating system vendors 
change their systems --- typically to harden them against attack --- in ways that can undermine code that has for decades been used to dynamic conversion of source code into machine code, to checkpointing generated code and reloading it in a later session not expecting the system address map to have changed much, and distributing versions of their software without the choice of either paying somebody to attach a digital signature (even though they may have no revenue stream to cover such costs) or to accept that those who download and try to use it may see warning messages to the effect that the code might well be dangerous. The skills and techniques involved in all of these (which apply to all currently active systems) feel very much within the scope we are discussing here.

A quite different style of porting is called for when some significant intellectual development has been made and the originator created an implementation of it. They will typically have done that as an extension to or as code in the context of some existing computer algebra system. They might well make their code available to all under some sort of generous and flexible license.
But those who would like to either test and explore the work or leverage it to enhance the capabilities of their own system may find they have something of a porting mountain to climb. We can cite two separate cases of this that we are interested in which both happen to involve indefinite integration. They also both happen to be cases where the originator of the code of interest is deceased and hence not in a position to assist:
\begin{enumerate}
    \item Manuel Bronstein worked on systematic schemes for symbolic integration (antiderivatives)~\cite{Bronstein}, building on the work on Risch\cite{Risch}. His implementation for \sys{Axiom} (see our section on \sys{Aldor}) is probably the most complete development of the mathematical basis for the task. However now not only is Bronstein not with us, but the major developer who has been working on that for many years appears to have given up. So there is a serious issue of portability that has to be overcome so that the practical aspects of Bronstein's work being lost forever. 
    An optimistic response to this is to consider AI transformation from the code expressed in \sys{Axiom}'s rather particular language into something more widely used;
    \item Albert Rich had been taking an almost diametrically opposite approach to indefinite integration, and had been developing a very extensive rule-based system called \sys{Rubi}\cite{RUBI}. Part of the desire there was not just to evaluate the integrals but to generate results in concise and tidy form --- something that schemes based on the formal mathematical approach often fail to achieve.  For example, in its most recent version, \url{https://12000.org} shows that given the integral $\int \frac{dx}{x^5(a+bx^8)}$ (test number 245), \sys{Mathematica}~\cite{Mathematica} produces output
    
    \begin{footnotesize}
    \begin{verbatim}
(-Sqrt[a] + Sqrt[b]*x^4*ArcTan[Cot[Pi/8] - (b^(1/8)*x*Csc[Pi/8])/a^(1/8)] 
+ Sqrt[b]*x^4*ArcTan[Cot[Pi/8] + (b^(1/8)*x*Csc[Pi/8])/a^(1/8)]
+ Sqrt[b]*x^4*ArcTan[(b^(1/8)*x*Sec[Pi/8])/a^(1/8) - Tan[Pi/8]]
- Sqrt[b]*x^4*ArcTan[(b^(1/8)*x*Sec[Pi/8])/a^(1/8) + Tan[Pi/8]])
/(4*a^(3/2)*x^4)
    \end{verbatim}
    \end{footnotesize}
    while \sys{Rubi} gives
    \begin{footnotesize}
    \begin{verbatim}
(-(1/(a*x^4)) - (Sqrt[b]*ArcTan[(Sqrt[b]*x^4)/Sqrt[a]])/a^(3/2))/4
    \end{verbatim}
    \end{footnotesize}
    Rich had coded his rule-set using \sys{Mathematica}, and using it in that system it performs remarkably well on various test suites. He made his rules freely available, but when one attempts to use them free-standing it gradually becomes clear that, at least in the form that he left them in, there is much reliance not just on \sys{Mathematica} syntax (which is rather easy to reproduce) but on the detailed semantics of the pattern matching there and the incidental simplification that \sys{Rubi} expects \sys{Mathematica} to do for it. The code there is still available and can be useful for those who have a \sys{Mathematica} license (or for that company to embed within their commercial system), but porting it out for use elsewhere has proved non-trivial\cite{NormanJeffrey}.   
\end{enumerate}

When significantly mathematical code is to be ported then there are a wide range of ways in which different systems interpret similar notations in different ways. This can include the treatment of multi-valued functions and indeed whether formulae are to be viewed as existing in the real or complex domain or some more complicated world. In terms of communication with users $x \sqrt x$ and $x^{3/2}$ may alter the appearance of formulae enough to be practically important even if mathematically fairly trivial. Use of mathematical notation is notably flexible and in different contexts conflicting conventions can apply. As a simple concrete instance, the Bernoulli number with index 1 may be either $+1/2$ or $-1/2$ depending on the use to be made of it! We elaborate on some more of these challenges in Section~\ref{sec:portable-math}.
\subsection*{Why portability?}
The very essence of the scientific method is that when an advance has been
made others should be able to reproduce and validate it. In computer
algebra's early days the papers presenting new algorithms showed techniques
that the reader could fairly readily re-implement, but even then a number of them included
the code that the authors had produced. These days, when code is usually way too complicated to recreate from a mere description, it is commonplace for
code to be deposited on some publicly available archive. If that code will
only work using exactly the computer configuration and support software
context that the author enjoyed then even in the short term its usefulness
to others may be constrained, and very typically over time as that setup becomes
out of date it will lose value. In the case of a package to perform some
algebraic transformation in a way that moves the state of the art forward
there can be a level of ambiguity as between the new work and the strengths
of the platform it is built on, so other workers could reasonably want to try
it atop a different base. The issues of being able to do this alongside the
ones forced by continuing changes in the world of computers puts portability
at the heart of all we do. The authors here have perhaps been particularly been
brought face to face with this issue by the examples regarding indefinite
integration cited later on, where in one case their own earlier contribution
had included integration code that needs to be updated or replaced and in the other where there was direct involvement in the creation of the programming system
used in one of the projects considered -- but where the whole language now looks at risk.

Sometimes making software portable adds a non-trivial burden to the
implementer, whether it be a 1960's \sys{Fortran} programmer needing to know
about ``definition at the second level''\cite{secondlevel} or today's \Cpp{} users about
``strict aliasing''~\cite{strictaliasing} and many other easy to use idioms that can lead to
unexpected behaviour on some, but not all, platforms. There can also be coding patterns
that may lead to widely varying system-dependent performance (\sys{Verilog} can
exhibit this in spades, but those who have worked in high performance computing will also be very aware that tuning can lead to dramatic performance shifts). However, for many people whose career paths depend mainly on their publication record, the incentive can be to get an
implementation of their ideas working well enough on just the computer they
happen to be using this month, report on results and timings and move on
--- leaving their code available to all to download but without true
certainty that it will remain usable either elsewhere or in the future.

So there are perhaps three classes of people who worry about portability:
\begin{enumerate}
    \item Those building an artefact 
    that they hope many people will use and find so central to their process that they cite its originator, leading to the distribution of the software being viewed as having high impact. Both the early work on \sys{Reduce} and \sys{Maple} took this stance;
    \item Employees in a company preparing commercial software where portability across existing platforms and to potential future ones may broaden the market. \sys{Maple} and \sys{Mathematica} must have, of necessity, followed this pattern;
    \item Developers of open source software that manages to attach attention and get used, because the users may be distributed across a wide range of computing environments and will report when the software fails to behave for them. Those currently curating \sys{Reduce} are grateful for the early portability emphasis and some of them could be viewed as fanatics in that regard to the current time.
\end{enumerate}
We fall into the first category when we find that, against plausibility, software that we developed many years ago is still in use, and we look back in horror at some of the coding styles we originally used and but feel that keeping it current extends the pleasure we get from knowing that our code is used to solve real problems. But more than that we are in the third category where we take pride in keeping the software we support going on older systems but also arranging it can take advantage of whatever the latest platform it. In that context
we also believe that if open source software is to attract a new generation of maintainers, the code should look tidy (and ideally this associates with portability) and is not limited by only running on archaic platforms.  Documentation can be helpful in principle, but is often out of date and so misleading. Tim Daly had been re-working \sys{Axiom} to put much of its code in ``literate'' form~\cite{LiterateProgramming} with the explicit objective of making it longer lasting: unfortunately this failed to lead to a big influx of new collaborators---it illustrates some of the challenges we
are commenting on here.

Making code portable to begin with and keeping it able to run in a changing world is not always glamorous and and if done well will be almost invisible --- so we believe that there are not a huge number of presentations at academic gatherings outside those that specialise in software engineering that report on it. We hope here to in some small way address the
imbalance where the total work going into invisible (but challenging) system maintenance may be very comparable with the amount that goes into developing a new and flashy specialist extension. That will then need long term support if it is to truly deliver on its promise!

\section{Early Portability in Computer Algebra}
\label{sec:earlyportability}
\subsection{1960s: Assembly and Fortran}
Some of the earliest computer algebra software packages were written in system-specific assembly code, either as stand-alone programs or as libraries to be used from a Fortran program.   
\sys{Alpak}\cite{Alpak} was one such library package.  

\sys{Alpak} was rewritten as a set of \sys{Fortran} subroutines and paired with a language translator to become \sys{Altran}\cite{Altran} whose portability is mentioned by Johnson and Richie\cite{UnixPortability} when they were considering 
other software portability issues. 

\sys{Altran} programs looked like Fortran with the added \texttt{algebraic} type.
There were a number of \sys{Fortran} implementations from different vendors at the time, with incompatibilities, 
\sys{Fortran} implementations had many variants and we understand that these were accommodated using the \sys{M6}\cite{M6} macro processor.  
Later, the \sys{Fortran 66} language standard was produced, which in principle allowed \sys{Fortran}-based software to run across a variety of hardware platforms.

The \sys{Camal}\cite{Fitch09} computer algebra system, developed by David Barton, Steve Bourne and John Fitch, was originally written in assembly code for the Titan  computer at Cambridge. There were just two very similar computers in the world, and another three that were fairly close relatives.

Over the same period, George E. Collins was developing a set of libraries, \sys{PM} \cite{CollinsLibrary}, for big integer and polynomial arithmetic, first at IBM, then at the University of Wisconsin.  We are under the impression that these began as assembly code libraries before the development shifted to Fortran, but would appreciate confirmation of this.

\subsection{1970s: \sys{Lisp} and \sys{BCPL}}

As the 1970s commenced, it became evident that it was undesirable to be rewriting software in different assembly languages for successions of machines.   Writing in \sys{Fortran} provided portability, but dynamic storage allocation was difficult in this setting.

Two alternative approaches emerged.   The first was to rely on \sys{Lisp}\cite{Lisp} as an implementation language, as \sys{Lisp} had the concept of dynamic data structures at its core.   The second was to rely on lower-level ``systems programming languages'' to provide efficient resource management. At the start this meant \sys{BCPL}\cite{BCPL}, although in due course \sys{C}\cite{C} claimed the dominant place in that ``ecological'' niche.

From its beginnings at the end of the 1950s, \sys{Lisp} had matured and taken its place as a recognised family of closely related languages.   There was no notion of a single, standard \sys{Lisp} language.  Important variants around 1970 were \sys{MacLisp}\cite{MacLisp}, from MIT, and \sys{InterLisp}\cite{InterLisp}, then from Bolt, Beranek and Newman. For some time, both of those had a clear enough common core that migrating software between them was reasonably sane for quite a while. 

However by the end of the 1960s dialect divergence was sufficient to cause trouble and, Anthony Hearn introduced \sys{Standard Lisp}\cite{StandardLisp1} to serve as the foundation for the \sys{Reduce} system. This was not so much intended for adoption outside \sys{Reduce} usage, but it defined the expectations that \sys{Reduce} had. It was consciously conservative and a bit minimal, and hence it was not especially attractive to those who wanted to develop and extend \sys{Lisp} for direct use. So meanwhile
the \sys{Macsyma} system was built on \sys{MacLisp}, first led by William Martin and then Joel Moses from 1968 to 1982 at MIT, and the \sys{Lisp} dialect used there collected many many add-ons, where Joel Moses characterised the language as ``a ball of mud''\cite{ballofmud} that one could add arbitrary extras to it without altering its already slightly odd outline.  Later, a port of \sys{Macsyma} to the VAX architecture was achieved at UC Berkeley by creating a \sys{MacLisp}-like base, \sys{Franz Lisp}, by Richard Fateman and his students.

The \sys{Scratchpad} system was built on \sys{Lisp/VM}\cite{LispVM}, led by Jim Griesmer at IBM.  The IBM \sys{Lisp} system was led by Fred Blair and later Cyril Alberga. (Note, this \sys{Scratchpad} was retrospectively named \sys{Scratchpad I} to distinguish it from the later, distinct system \sys{Scratchpad II} which became \sys{Axiom}.) \sys{Scratchpad} (I) was seriously notable from the portability perspective because it started life by amalgamating major components from almost all of the \sys{Lisp}-coded algebra projects extant around 1970, when the project was in its early stages and the various \sys{Lisp} dialects were still close enough that code for one could be adapted for use on another with only a reasonable amount of effort.

In each of these cases, the \sys{Lisp} system isolated the algebra system from the details of the supporting hardware, and storage management, system checkpointing and file handling issues were handled with it --- with extended precision integer arithmetic soon joining the collection of expected capabilities.

The different \sys{Lisp} systems were not equally available on all platforms, so some systems made an attempt to run on more than one \sys{Lisp}, despite the differences.   \sys{Reduce} is particularly notable in this regard.    A version of it dated 1973 has been recovered and it can still be run --- and amazingly much of the code in it can be recognised in the most recent versions of the system. This historical relic is available for download on the main \sys{Reduce} website at \url{https://sourceforge.net/projects/reduce-algebra}.  The strategy there was to rely on a minimal set of core \sys{Lisp} functionality, and to reproduce necessary higher-level function as a set of \sys{Lisp} functions relying only on that core.   A similar approach was later adopted for \sys{Axiom}.

The usual approach to building these \sys{Lisp} systems was to have an assembly code core written in a ``LAP'' (assembly code in \sys{Lisp} syntax), and for most of the system to be written in \sys{Lisp} itself.   Some \sys{Lisp}s were written in other languages, including \sys{Cambridge Lisp}\cite{CambridgeLisp}, which was written in \sys{BCPL}.

\sys{BCPL}\cite{BCPL} was designed by Martin Richards of Cambridge University in the second half of the 1960s as a simplified form of the \sys{CPL}\cite{CPL} language designed earlier in the decade.   It was designed to have a close mapping to hardware and served as a low-level ``systems'' programming language,  with a particular objective that it support the implementation of the full \sys{CPL} compiler. It also represented a landmark in having a compiler that could be portable and that was easy to retarget for other systems. This led to it being widely supported and in due course to it being a precusor of \sys{C}.   When changes in Cambridge computers made a machine code version of \sys{Camal} inappropriate, the concepts were ported even though the code could not be, and it was was rewritten in \sys{Algol 68} and soon after that in \sys{BCPL}.

\subsection{1980s and beyond: \sys{Lisp} and C}

At the end of the 1960s, the \sys{B} programming language was developed at Bell Labs, derived from \sys{BCPL}, by Ken Thompson and Dennis Richie.   The first version of \sys{Maple} was written in \sys{B} in 1980 as this was the systems programming language available on the Honeywell system upon the which the \sys{Maple} kernel was first developed.   

The programming language landscape was continuing to evolve with several closely related languages available and no clear leader evident.  Among the alternatives were \sys{B}, its follow-on \sys{C} and a related University of Waterloo language, \sys{Port}.  As these had overlapping syntax and concepts, it was possible to write the \sys{Maple} kernel in the intersection and use a bespoke macro-processing language (\sys{Margay}) to provide \sys{B}, \sys{C} and \sys{Port} variants.   Most of the \sys{Maple} system was written in its own bespoke programming language. One can see that as paralleling the way in which \sys{Reduce} had avoided a total commitment to one particular platform, though it was clearly even broader in scope.

In the mid 1980s, the development of \sys{Scratchpad II}\cite{ScratchpadII} was underway.   Its implementation used a layered set of programming languages.   
At the core was the \sys{Lisp/VM} system, itself built on \sys{LAP} assembly code and \sys{Lisp} bootstrap files.
The next layer, including most of the interactive system and first compiler, was written in \sys{Boot}, a syntactic sugaring of \sys{Lisp} to an \sys{Algol}-like syntax and with destructuring assignments and other features that its authors expected to be particularly relevant while constructing a symbol
manipulating program.

\section{Issues Encountered in \sys{Lisp} Evolution}
\label{sec:lispevolve}

We now give a personal account of issues encountered by one of us (ACN) in developing a \sys{Lisp} base for the \sys{Reduce} system:

The first \sys{Lisp} ACN and John Fitch implemented so they could run \sys{Reduce} was
``\sys{Cambridge Lisp}'' and was coded in \sys{BCPL}. The data representation used high
bits in a word to tag things (and the IBM mainframe in use used 24-bit
addressing so the top 8 bits were free). For that the bignums used $10^9$
as a base because 
doing that made only a linear change in
the cost of arithmetic but made reading and printing linear not quadratic.
Numbers were never going to be big enough to mess with $n \log n$ schemes!

After a significant while it became clear that \sys{BCPL} was becoming less
central to the world and that byte- not word-addressed machines were 
very dominant. So \sys{CSL} (\sys{Standard Lisp} was coded in \sys{C}).
By then \sys{C} was
\sys{Ansi C} so it was not as nasty as \sys{K\&R C} had been. Bignums were made out of
31-bit chunks because \sys{C} did not give trivial access to the carry bit.
Data was tagged in the low 3 bits. Essentially small integers were represented as immediate values, cons-cells had a tag dedicated to mark
references to them and all other data used a header word containing
further discrimination as well as a size field.

Again the years passed. 64-bit machines arrived so it was necessary to review all the
code for 32-bit sensitivity. Bignums remained at 31-bit digits. The thing
that was most amazing was the checkpointing scheme. A system image was
checkpointed by writing a simple binary dump of the heap image and a word
giving its start address. On reloading, the heap was parsed well enough to
relocate and, for 64-bit use, it was possible to reload and map either way between 32
or 64-bit images. One might imagine that this was exploiting \sys{C} with a
detailed belief about storage representation (it also allowed for
byte-sex [bit order], so a heap image made anywhere could be used anywhere else).

Yet more years passed and ACN concluded that if eventually anybody else was going to
maintain this software it needed to be in \Cpp{} not \sys{C}.  The \Cpp{} language is much more fussy about ``\verb+const char*+'' so
there was a deal of work there. It was now possible to use ``\verb+//+'' comments
officially rather than ``\verb+/* .. */+'' ones that were all the older \sys{C} systems
certified as legal (whatever \sys{gcc} actually coped with).

Since then, the implementation has gradually moved to use more \Cpp{} facilities. But some of the more radical changes may be worth noting:
\begin{itemize}
\item The allocator and garbage collector were ripped out
and replaced by a new conservative one.  On 64-bit architectures, conservative collectors are less problematic --- falsely identifying a data word as a pointer (and thus retaining garbage) is far more probable on 32-bit architectures.
\item A new arithmetic package has been written that
will use 64-bit digits for bignums and which will be faster then the old one as well as cleaner code --- but despite quite a few years in
gestation that is not the standard one in place yet.
\item The checkpoint/restart scheme has been replaced with one using techniques akin to data serialization.
\end{itemize}
A point to be made about these and the many other changes that have been made at that level, is that they are isolated from the code that performs algebra as such and that they were motivated by the need to support \sys{Reduce} on a broad range of modern systems and with the best possible
performance and the best changes for future portability and maintenance.

For \sys{Reduce} one can explain to people that when it was young
you just said ``it is coded in \sys{Lisp}'' and they thought they understood. But
a better way to explain it now is that it has its own language and the
parse trees for that and the raw data that is worked on happen to look
very much like Stanford \sys{Lisp 1.6} --- but that it is an internal abstraction layer and
that almost nobody extending or maintaining \sys{Reduce} works in \sys{Lisp} as a source
language. So the ``\sys{Lisp} implementation'' used stands a good chance of really
being the ``\sys{Reduce} kernel'' and not being used for any \sys{Lisp} activity
elsewhere. And the main versions used are thus customised and tuned to
support \sys{Reduce}. Hearn and his collaborators also found it helpful to update and extend the specification of the \sys{Lisp} standard\cite{standardlisp2} that they would rely on, so the variants cited here agreed to that new contract. ``\sys{CSL}''\cite{CSL} is coded in \Cpp{}, while ``\sys{PSL}''\cite{PSL} builds on the Utah \sys{Portable Lisp Compiler}\cite{PortableLispCompiler}  and is coded in ``\sys{syslisp}'' which
is a \sys{Lisp} that has functions like \verb+(wplus a 1)+ that increments the machine
representation of the value of a, and so that and its friends let it
implement bignums and garbage collection rather as if it was a parse tree
for \sys{C} where ``\verb$int c = a+b$'' has turned into ``\texttt{(prog (c) (setq c (wplus a
b))...}''

It may also be asserted that the \sys{Reduce} language is the successful version of \sys{Lisp
2}\cite{Lisp2}. McCarthy thought in terms of S-notation and M-notation and \sys{Lisp 2} was
to support the latter (but the project basically failed). \sys{Lisp 2}
``had largely \sys{Lisp}-like semantics and ALGOL 60-like syntax''. The \sys{Reduce}
``rlisp'' notation is a really rather close relative of what the \sys{Lisp 2}
people designed --- but it was made to work, work portably and has survived
in use for very many years. One can see \sys{Scratchpad}'s \sys{Boot} as very much a  variant on the same concept.

Note that \sys{Reduce} and \sys{C} were being invented at about the same times, but 
it took longer for \sys{C} to be distributed beyond Bell Labs than for
\sys{Reduce} to get into circulation.

\section{Case Study: \sys{Reduce}}
\label{sec:case-reduce}
The code that eventually became known as \sys{Reduce} was originally coded directly in \sys{Lisp 1.5} --- but by the end of the 1960s it had been reworked so that it was written in its own ``Algol-like'' syntax that for the system programmer offered \sys{Lisp}-like semantics.
For example, the following computes Legendre polynomials:
\begin{verbatim}
procedure factorial n;
  for i := 1:n product i;

for k := 0:10 do
  write df((x^2-1)^k, x, k)/2^k/factorial k;
\end{verbatim}
Building the system called for a fairly modest size parser in regular \sys{Lisp} to bootstrap things until the system's better and more complete version of the same parser expressed in its own language could be installed. This scheme provided significant isolation from dialect variations between competing \sys{Lisp} implementations --- the bootstrap parser could need to be adapted to allow for differences in support for character handling, but then the transformation from textual source code to internal S-expressions readily provided a location where adjustments to the use of \sys{Lisp} functions could be made. To give a concrete example of the sort of adjustment needed, we note that the \sys{Lisp} operators \verb.map. and \verb.mapcar. that apply a given function to each component of a list took their arguments in differing orders basically depending on whether the \sys{Lisp} implementation concerned followed East Coast or West Coast traditions. \sys{Reduce} could adapt for that by including conversion code as part of its parsing phase.

Over time and as the number of computers (and hence \sys{Lisp} systems) that people wanted to run \sys{Reduce} on grew this scheme which amounted to source to source translation from the dialect \sys{Reduce} expected to whatever was locally available became clumsier and more onerous. In response to this \sys{Reduce} codified a \sys{Lisp} dialect it would use (calling it ``\sys{Standard Lisp}''\cite{StandardLisp1}) so that host systems could arrange to provide a compatibility layer that exposed this agreed version, and the code within \sys{Reduce} being written based on that could then just run without need for further change. To make this process as easy as possible \sys{Standard Lisp} mandated only facilities it expected that all plausible \sys{Lisp}s would be able to support, so it did not provide any of the more elaborate additions to \sys{Lisp} that those who were using the language directly were developing. For instance more or less elaborate loop constructs were present in the \sys{Reduce} source language but got expanded to simple forms by the time a \sys{Lisp} ``parse tree`` had been generated, so there was no need for the supporting \sys{Lisp} to provide clever iteration functions.

So the increasing bulk of \sys{Reduce} being coded in its own language and sitting on a rather minimal base really helped with portability, but two challenges
could arise. One was that newer parts of the system sometimes needed facilities not captured in Standard \sys{Lisp}. Perhaps the best example to quote would be modular arithmetic. This could be provided by merely defining all the necessary functions in terms of ordinary arithmetic with a lot of calls to the remainder function, but really it would be better if implemented at a lower level. Well \sys{Reduce} could cope with that! It could include portable reference implementations of everything is needed, but a particular implementation could offer built-in and presumably better versions and then tag the relevant function names 
to 
cause \sys{Reduce} to skip instating its reference code. This strategy maintained the policy that the bulk of the code should only very very rarely need to be conditialised based on the platform it was running on.

The second challenge was of course performance. Main-stream \sys{Lisp}s were increasingly focusing their implementation efforts on supporting the full richness of environment expected by users who were coding directly in the parenthesised notation and the optimisations they were delivering were not necessarily the ones most important to \sys{Reduce}. So as a spin-off activity \sys{Reduce} first developed its own explicitly portable \sys{Lisp} compiler, and went on from there is develop of complete \sys{Lisp} (PSL\cite{PSL}) explicitly providing (a superset of) \sys{Standard Lisp}, reasonably easy to move to new machines because it used the portable compiler, and particularly tuned to run \sys{Reduce} well. PSL was essentially all coded in \sys{Lisp} as extended with functions that were there to compile into simple operations on machine registers and memory. In broadly the same time-frame \sys{Cambridge Lisp}\cite{CambridgeLisp} was implemented mostly in BCPL but also using a version of the \sys{Reduce} project's compiler, and BCPL's focus on portability meant that ended up being used on a range of the emerging small computers from Amiga, Atari, Acorn and the like. In each case these distilled portability away from the parts of \sys{Reduce} that supported algebra, and they could both focus on trying to be economical in memory and good in terms of speed for the particular load that \sys{Reduce} placed on them.

Somewhat later on the use of BCPL became less tenable as \sys{C} took over the ecological niche it had inhabited and as Unix became a dominant operating system in academia. So \sys{Cambridge Lisp} was replaced by \sys{CSL}\cite{CSL}, which was a fully fresh \sys{Lisp} this time coded in \sys{C}. A particular feature of this was that its almost entire purpose was to act as a kernel for \sys{Reduce} rather than as a \sys{Lisp} for more general use. At the time memory often limited problem solving ability. To save space \sys{CSL} compiled the \sys{Lisp} that came out of the \sys{Reduce} parser into a compact byte-coded form which it could then interpret. One initial insight was that the bytecode interpreter was likely to fit within the caches of the more important computers of the times, and that would lead to the overhead of interpretation being much less severe than one might have at first feared. The second performance trick was to profile a wide range of \sys{Reduce} applications and on that basis identify the most crucial functions. Those were then compiled from \sys{Lisp} into equivalent \sys{C} code that followed the internal \sys{CSL} conventions and got passed through the system \sys{C} compiler and linked in as parts of the main \sys{CSL} executable. By then the leading \sys{C} compilers (notably gcc) were starting to provide optimisation capabilities distinctly superior to the more generic ones present in the available \sys{Lisp} compilers.
The \sys{C} code in \sys{CSL} was kept fairly conservative with a view to keeping it as portable as was reasonably possible.

Because this story spans many years, one should not be surprised to notice that just as BCPL had started to look long in the tooth, at some stage the \sys{C} code in \sys{CSL} started to be reworked into \Cpp{}. The changes there forced further case regarding for instance the difference between immutable strings and updateable arrays of characters, the issues of ``strict aliasing'', and the fact that arithmetic overflow on signed integers was deemed undefined. Responding to each of these will have tended to enhance some mix of reliability and portability of the code-base. One of the more concerning issues while that transition was going on was the proper delay time between \Cpp{} standards codifying useful new features and the moment that assuming their availability would not seriously compromise attempts to build the code on older machines. While it is often reasonable to expect that personal computers will be kept well up to date with the most recent versions of compilers and other software, major institutional systems can prioritise stability to an extent that to a developer who has not encountered that before it at first seems that they have hardly moved on from the ark. \sys{CSL} has tried to steer clear of reliance on new language features until they have had plenty of time to propagate to all potential use-sites. This is again in line with the long-standing \sys{Reduce} philosophy of preferring easy portability over extra and glossy features. Over its lifetime, the storage allocation and garbage collection parts of \sys{CSL} have been re-written several times, the scheme for preserving a system on disc for later re-loading has been changed also several times and the long integer arithmetic implementation altered  rather substantially. But from the outside, all these changes tried to be invisible so that the \sys{Reduce} system sitting on top did not need to pay any attention, but could benefit from speed improvements or reliability improvements.

\section{Case Study: \sys{Aldor}}
\label{sec:case-aldor}
\sys{Aldor} is a programming language for mathematical computation
created as an extension language for the \sys{Axiom} computer algebra system developed at IBM Research.  It grew out of the un-named language described in~\cite{JeTr81}, invented to express mathematical algorithms generically with algebraic restrictions on parameter types.
For context, this was a decade before \Cpp{} had templates and some three decades before it introduced ``concepts'' to qualify parameter types.

\sys{Aldor} purposefully retained most of the ideas and syntax of~\cite{JeTr81} while generalising the type system and providing features to allow clear separation of base language and libraries. 
Functions and types were first-class values and
the language design allowed efficient compilation.
Dependent types provided genericity and 
the structures of modern algebra (such as rings, fields and modules) can be represented naturally as type categories. 
An optimising compiler for \sys{Aldor} could generate \sys{C} code in the \sys{Ansi} (\sys{C90}) or \sys{K\&R C} dialects, for stand-alone or linked programs, or \sys{Lisp} in the \sys{Common Lisp}, \sys{Standard Lisp} or \sys{Scheme} dialects, for embedding in systems based on those.
For a fuller description, see~\cite{Aldor-Issac,Aldor-CAH,Aldor-UG}.
Recall that during its development \sys{Axiom} had been named \sys{Scratchpad II}, and similarly \sys{Aldor} was named $A^\sharp$ (and briefly as \sys{Axiom-XL}).

The main development of \sys{Aldor} spanned a period of about ten years, from 1984 to 1995.
It began as an effort to produce an improved implementation of the language of~\cite{JeTr81}.  It was written in the ``\sys{Boot}'' language, described earlier, and it translated algebra code to \sys{Lisp} to run within the \sys{Scratchpad II} system. 

In 1990 it was decided to end this first effort and begin afresh with a \sys{C}-based implementation. 
There were several factors at the time that informed this decision:  The \sys{C} programming language had become widely used and was gaining momentum, while the opposite was true of \sys{Lisp}.  Therefore competency in \sys{C} programming was more widely available for a development team.  This development was taking place in the heyday of platform diversity and \sys{C} implementations were available for almost all potentially relevant platforms.   Importantly, although carefully written \sys{Lisp} code could sometimes reach the efficiency of \sys{C} code, being able to generate highly efficient code remained challenging in a \sys{Lisp}-only environment. Finally, being written in \sys{C} would allow the compiler and generated code to be integrated into a wide variety of settings, either directly or via readily available foreign function interfaces.

We now detail how the \sys{Aldor} compiler was structured to achieve portability across the wide range of platforms while keeping a uniform, clean code base.  
Bear in mind that, at the time,
operating system interfaces and architectures varied greatly and
C language and library implementations had rather hit-and-miss adherence to standards.

A wide range of \emph{operating systems} were supported, including
 Microsoft DOS and Windows 3.1, IBM OS2 and CMS, Digital Equipment VAX VMS, Apple MAC System 7, and a wide range of Unix variants including AIX PS2, AIX RT, IX RS, AIX 370, AIX ESA (all from IBM), SunOS and Solaris from Sun Microsystems, the Open Software Foundation's OSF/1, as well as Unicos, Irix, Next, Linux, Ultrix and HP UX.
The \emph{C environments} supported included those from
Borland, Metaware, Metroworks, Microsoft, the Free Software Foundation (\sys{gcc}), four from IBM, DEC's VMS and Ultrix \sys{C}, as well as those from Apple, Cray, Hewlett Packard, MIPS and Sun.
The \emph{architectures} supported included Intel 8086 and 80386, Motorola 68000, PowerPC, IBM 370, RT and RS, DEC VAX and AXP, Sun Sparc, MIPS, Cray and Hewlett Packard PA.

We see that there were a dizzying array of compilers, operating systems and architectures, each with their own quirks.  There were multiple representations for floating point numbers, pointers (segmented, \textit{vs} flat address space, and with varying addressability), and quite different notions of file systems.

To deal with this complexity, the compiler software 
was organised in a set of layers, providing successively higher levels of function, and each layer relying only on itself and lower layers:
\begin{itemize}
\item Layer 0: Portability
\item Layer 1: General library
\item Layer 2: Compiler-specific data structures
\item Layer 3: Compiler phases.
\end{itemize}
In this article, we are primarily concerned with the portability layer and, to a lesser extent, the general library layer.

The portability layer provided a set of very low-cost abstractions to give a common coding interface for all the supported platforms.
This allowed efficient low-level programming while shielding the main body of the code from defects in the various \sys{C} compilers (of which there were many) and providing common concepts where operating systems differed.   
This  approach necessitated a certain amount of system-specific implementation of low-level functions (for a recent version, 6829 lines of \sys{C} code for all the operating system variants listed above out of a total of 199,939 lines).
This portability layer was the \textbf{only} layer in which platform-specific \verb+#ifdef+s could occur, and there were about 370 of these.

It was assumed that any part of of the \sys{C} library used would behave according to the language standard, even though various platforms had missing parts, varying return types and other inconsistencies.
This was achieved by having the portability layer provide a wrapper for each standard library header file, \textit{e.g.} code would use \verb+#include "stdio.h0"+ instead of \verb+<stdio.h>+. 
These wrappers would do whatever was necessary to adjust the available libraries to behave according to the standard.  

Within all but a strict subset of the portability layer, specific compilers, operating systems and architectures were not be used to conditionalise code.  
Instead each environment could activate certain named properties, such as \spacetune{\linebreak}
\verb+CC_noncanonical_pointer+
or
\verb+CC_broken_toupper+.
All told, there were about 50 of these, and they were used to configure the standard library wrapper files.

The portability layer also provided a range
of macros to encapsulate code that might have different variants depending on the named properties.
For example, \verb+ptrDiff(p,q)+ had different implementations depending on whether or not pointers were canonical.  Strictly speaking, the \sys{C} standard would not allow subtraction of pointers into two different objects, so the \verb+ptrDiff+ operation was needed to support garbage collection.
Another example would be that in some (but not all) environments, functions to be used as signal handlers had to have a special modifier in their declaration, and this modifier varied.  So a macro \verb+SignalModifier+ would always be used in such declarations, and it would sometimes expand to a keyword and otherwise to nothing.  All told, there were about 80 of these macros.

Although the \sys{C} standard~\cite{Std-C90}
  was very new and support for it was lacking on many important platforms, the \sys{Aldor} source code
 would be written to the standard.  A software tool was written to convert function prototypes and new-style function headers if needed on a given platform,
 \textit{e.g.} 
 \verb+extern int f(char *s);+ and \verb+int f(char *s) { .. }+,
 to \sys{K\&R C} style,
 \textit{e.g.} 
 \verb+extern int f();+ and\spacetune{\linebreak} 
 \verb+int f(s) char *s; { ... }+.

The portability layer also included an abstraction of the operating interface.  This included
\begin{itemize}
    \item hooks to run at the start and end of a program (\textit{e.g.} for license management),
    \item message display (\textit{e.g.} for a GUI or command line),
    \item spawning and running subprograms and directing their input and output,
    \item measuring time,
    \item obtaining and adding values to an ``environment'',
    \item interacting with the file system, including file name manipulation, input and output, file hierarchies (\textit{e.g.} directory tree structure, separate disks, \textit{etc}), detecting existence and permissions, temporary files, and so on,
    \item handling signals, including breaks and faults,
    \item and memory management.
\end{itemize}
Most of these were provided with straightforward, operating system-specific code.
The last item, memory management, deserves special mention.

A design decision for the language was that \sys{Aldor} code should be able to participate as widely as \sys{C} code in foreign function interfaces.
This meant that pointers to \sys{Aldor} objects could be
passed to foreign functions, and pointers to
foreign objects could be passed to \sys{Aldor} functions.
As these pointers could be captured, it was therefore necessary for the \sys{Aldor} run time to be able to detect objects that were live and immovable in the \sys{Aldor} heap by virtue of being pointed to by foreign memory.  To support this, a single function, \texttt{osMemMap}, was provided to provide information about the layout of memory.
It returned a vector of structures describing the regions of memory accessible to the running program.  Each region was classified as one of initial data, dynamic data or stack.  
There was
no assumption that there was only one stack, nor that stacks were all in one region.  
The calling program could identify foreign memory by keeping track of its own memory use.

\sys{Aldor}'s run time library provided a conservative garbage collector that made use of the \texttt{osMemMap} function.
The root set for garbage collection was obtained by
scanning all data regions outside the \sys{Aldor} heap to find all potential pointers into the \sys{Aldor} heap, including those from other memory managers.

By these methods, all the code above the portability layer was to be free of configuration \verb+#ifdef+s.
Beyond the portability layer, the general library layer provided enhancements to the standard \sys{C} environment to aid building a large project.
It provided a set of data structures, including a type-safe list template type, dynamically scoped (a.k.a. fluid) variables, arbitrary precision integer arithmetic, storage management with garbage collection, and conversion between native and {\sc Ieee} 754 floating point representations and between native and \sys{Ascii} charcters (\sys{Unicode} lay in the future).

To give an idea of the amount of portable \textit{versus} platform-specific code, we note the compiler and run-time system consisted of about 200,000 lines of \sys{C} code, 
1000 lines of 2-level grammar code, 
and 2,400 lines of \sys{Aldor} code.      
Of the \sys{C} code, about 8000 lines were devoted to the portability layer.  
About 2000 lines of \sys{C} code were for software tools were used to build the compiler. 
These included \verb+oldc+ (to convert  \sys{Ansi C} to \sys{K\&R C}), \sys{zacc} (to convert 2-level grammar to \sys{yacc}), and \sys{atinlay} (to convert the message database to pig latin, for multilingual testing).
 
Libraries in the \sys{Aldor} language completed the system, 
with  about 11,000 lines providing basic types and functionality 
and an additional 41,000 lines providing symbolic mathematical functions, including the $\sum^{IT}$ library~\cite{SumIT}.
This provided efficient algorithms for sparse and dense linear algebra, polynomial algebra (including factorization, resultants, etc), rational functions and series, as well as difference and differential operators over integers, finite fields and rational numbers.

The system design also considered portability of
the generated code.
As stated above, the \sys{Aldor} compiler generated \sys{C} or \sys{Lisp} code in various dialects.  To provide portability of the generated code separate files of \sys{C} or \sys{Lisp} macros were provided.

What we take from this experience is that,
with due care and attention to detail,
it has been possible to build a highly portable software system for computer algebra that runs
across a wide range of diverse platforms
by providing a low-level portability layer comprising about 3\% of the total code base.
The resulting system was able to produce code 
that ran at speeds comparable to hand-coded \sys{C} 
or \Cpp{}
for the sorts of problems these systems could handle~\cite{Aldor-Issac,SciGMark}.
As we discussed the topics covered here, we were struck by the parallels between the \sys{Reduce} and \sys{Aldor} experiences.

\section{The Role of Standards}
\label{sec:standards}
The issues raised in the case studies of
Sections~\ref{sec:case-reduce} and~\ref{sec:case-aldor} allude to the value of software standards.
In recent years, we have had mature, relevant standards upon which new computer algebra software systems can rely.
Principal among these are
\begin{itemize}
    \item the {\sc Ieee} 754 binary representation for floating point numbers~\cite{Std-Ieee-Floats},
    \item the \sys{Unicode} standard for character representation~\cite{Std-Unicode},
    \item the \sys{Common Lisp} language standard~\cite{Std-CommonLisp},
    \item the \sys{C} and \Cpp{} language standards~\cite{Std-C90,Std-C24,Std-Cpp17,Std-Cpp20,Std-Cpp23}, 
    \item the \sys{Posix} operating system interface standard~\cite{POSIX}, and
    \item the \sys{XML}, \sys{MathML} and \sys{OpenMath} standards~\cite{Std-XML,Std-MathML,Std-OpenMath,OpenMath97}.
\end{itemize}
Together, these address most of the issues raised in the portability of earlier systems, with the exception of low-level abstractions to allow garbage collection.

We have seen, particularly in Section~\ref{sec:case-aldor}, how coding to 
a standard benefits the elegance of a large code base, even when the standards are not actually met by the base platform.    
Providing a ``fix up'' layer allows the main code body to remain for the most part unchanged as the
platforms become more standard-compliant over time.

In this spirit, the \sys{CSL} implementation of \sys{Lisp},
has provided all of the short floating point arithmetic required by \sys{Common Lisp}.  The revision of \sys{Aldor}, currently underway, is based on the \Cpp20 standard, even though support for it is still only partial in the most popular \Cpp{} implementations.

Today, we have an embarrassment of available computing power as compared to the early days of computer algebra.  
We have had a convergence of platform variants so that the support of
\begin{center}
64-bit words + 8-bit bytes + byte addressability +
\\
\sys{Posix} + \sys{Unicode} + {\sc Ieee} 754 floating point + 
\sys{gcc}/\sys{clang}
\end{center}
covers almost all relevant modern environments. 
For other environments, such as antique systems or modern microcontrollers, being able to generate code for them on a modern architecture is a reasonable approach.

This common current platform obviates prior concerns relating to 
\begin{itemize}
\item the sizes for native integer arithmetic (16, 18, 32, 36, 64 bits) and consequently the transition to big integers,
\item the size and representation of floating
point numbers (IBM, VAX, Cray, {\sc Ieee} 754) and consequently the transition to big floats,
\item the size and representation of pointers
(\textit{e.g.} 16, 18, 20, 32, 36, ... bits, segment-offset pair \textit{versus} flat address space, \textit{etc}),
\item character sets (\sys{Ascii}, \sys{Ebcdic}, code-pages, \sys{Common Lisp} attributes, \sys{Unicode}),
\item byte ordering,
\item memory addressability (byte adressability, word adressability of various sizes, bit adressability [\textit{e.g} CDC Cyber 200]),
\item
Pathalogical machines with saturating overflow or traps on more things that are even slightly dodgy
\end{itemize}
One consequence of the reasonable assumptions that can be made for modern platforms is that the run-time support for systems has some portable-in-practice options for inexpensive tagging data by
pointer encoding.  The two most common techniques involve placing information in the low-order or high-order bits of pointers.  If a system allocates dynamic memory in multiples of $2^n$ bytes, then pointers to dynamically allocated objects will always have zero as their lower $n$ bits.  This allows those bits to be used for other purposes, for example as type tags.
Most widely used systems today use only the lower 48 bits of 64 bit pointers for virtual addresses, replicating the 48th bit for the higher bits. 
The adoption of {\sc Ieee} 754 floating point numbers allows the 48 bits of pointers to be encoded as NANs, with additional bits available for tags.\footnote{NAN-boxing has entered the mainstream. However, as larger virtual address spaces are supported, such as by Intel's 5-level page tables, this becomes more complicated.}

We anticipate that being standard compliant \textit{versus} having a finite number of low-level specialisations will remain a nuanced judgment call.
The issue of claimed \textit{versus} actual compliance of language implementations will remain
ongoing as standards are periodically revised.
For example, the \texttt{format} library was initially unavailable in the most popular \Cpp{} implementations compiling to the \Cpp20 standard.
Said another way, claimed standard compliance remains untrustworthy.

Looking to the future, we see \sys{Rust}, \sys{Typescript} and Web Assembly (\sys{WASM}) to be of rising importance.  These have not (yet) split into multiple implementations with serious incompatibilities.
Areas where new standards will be relevant to computer algebra are
\begin{itemize}
    \item GPGPU and NPU interfaces
    \item access to cloud storage (\textit{e.g.} \sys{DropBox}, \sys{GoogleDrive}, \textit{etc})
    \item access to web services (\textit{e.g.} the Online Encyclopedia of Integer Sequences~\cite{oeis},
    the L-Functions and Modular Forms Data Base~\cite{llmfdb},
    the Digital Library of Mathematical Functions~\cite{dlmf})
    \item libraries of formalised mathematical proofs (\textit{e.g.} \sys{Lean}'s \sys{mathlib}, \sys{Mizar}, \sys{Isabelle/HOL}), 
    \item interfaces to large language model services, 
    \item subscription- or fee-for-use- based licencing for information services.
\end{itemize}
As yet, these areas are not problematic, as there are not multiple variants in heavy use, but it is not unreasonable to anticipate that they may give rise to issues in time.

\section{Portability of Mathematics}
\label{sec:portable-math}
Now suppose that all of the platform portability issues were solved.  That would not be the end of the story.
From time immemorial, there have been disagreements around mathematical definitions,
and there remain so today.  In the literature different authors use different definitions for the same concepts.
Without agreement on definitions, it is not generally possible to take mathematical expressions computed by one system and use them in another.  
This applies both to the computer algebra world as well as the proof assistant world.

The portability of mathematics
itself remains an on-going open issue in the portability of computer algebra.
Differing definitions occur across mathematics' many fields.  We give a few examples where different authors take opposite stances:
\begin{itemize}
\item \textbf{Logic and foundations.} Do the natural numbers include zero?  Does the logic have the law of the excluded middle?
\item \textbf{Algebra.} Rings may be or may not be required to have a multiplicative identity.  Lie brackets may be defined as $[x, y] = u(xy - yx)$ with $u$ being $1$ or $i$.  Modules are sometimes implicitly taken to be left modules, but in noncommutative contexts the distinction between left and right multiplication matters.

\item \textbf{Algebraic geometry.}
Do divisors on a scheme follow the Weil or Cartier definition? These notions coincide on smooth varieties, but do not on singular varieties.

\item \textbf{Algebraic topology.} Poincar\'e duality is standard in oriented manifolds, but treatments differ in non-orientable cases.

\item \textbf{Homological algebra.} Boundary operators in simplicial homology can differ by sign conventions, also affecting chain complexes.

\item \textbf{Graph theory.} Do graphs allow loops, that is edges from a vertex to itself? Are complete graphs and edgeless graphs considered to be strongly regular?  This affects their classification and algebraic combinatorics.

\item \textbf{Complex functions.} Branch cuts can differ for multivalued functions.
For example the principal value of complex logarithm may be defined as $\log z = \log |z| + i \arg z$, where $\arg z \in (-\pi, \pi]$ or $[0, 2\pi)$, depending on convention.
Definitions of Fourier transforms vary on whether the forward or inverse transform has the negated exponent and how the forward and inverse transforms are normalised.

\item \textbf{Differential geometry.}
The Riemann curvature tensor differs by a sign depending on
convention. This affects derived quantities such as the Ricci
tensor and scalar curvature.
The Hodge star operator and volume forms depend on the metric signature and orientation.

\item \textbf{Formal proofs of the algorithms used.}
Some will view formal proofs as either not especially important or as
clearly too challenging to incorporate. Others take the view that without them the output from mathematical software has to be viewed with suspicion. One case where some implementation of \sys{Reduce} follows the path of caution is by using a library for double precision elementary functions that comes with formal proofs that the results are accurate and correctly rounded to the last bit. It does this perhaps more because this makes cross-platform regression test outputs match precisely than because it believes most of its users care about those low bits! At a higher level those who wish to see the algorithmic components of an algebra system proved properly will need to make their stances as regards the axiom of choice and the Riemann hypothesis clear, since various transformations rely on those.

\end{itemize}
Differences of definition such as these prevent mathematical data from being transported from one software system to another without careful attention.
We expect closer interaction between computer algebra and proof assistant systems in the future, so portability of formal proofs is relevant.
Differences in foundational logic as well as differences in definitions prevent the use of theorems from one proof assistant in another.  It has been an objective of the OpenMath~\cite{Std-OpenMath} and International Mathematical Knowledge Trust~\cite{imkt} to be able to capture these differences in a useful way.

A second form of impediment to the portability of mathematics is the encoding of mathematical algorithms.   
Significant tracts of mathematical knowledge are captured in the syntax of computer algebra and proof assistant systems.
Examples were given in Section~\ref{sec:meaning} of two approaches to symbolic integration locked in unfrequented systems.

\section{Final Thoughts and Conclusions}
\label{sec:conclusions}
The software portability challenges for computer algebra systems have changed over time.  
We have divided our attention roughly into four periods: 
\begin{itemize}
    \item the early platform period (prior to $\sim 1970$)
    \item the platform diversification period ($\sim 1970-\sim 2000$)
    \item the platform consolidation period ($\sim 2000-\sim 2015$)
    \item the mathematical consolidation period
    ($\sim 2015$ onward).
\end{itemize}

In the early platform period, there were very few available computer models that were sufficient to support non-trivial computer algebra applications, and system implementations could readily restrict themselves to the models within a given architecture family.

In the platform diversification period, the number of computing environments that could support computer algebra multiplied profusely.  
Computer architectures varied in memory models and data representations.
A plethora of operating systems emerged, each with its own abstractions for
interactive input,
memory management,
and file system interface.
At the same time, while a robust set of programming language implementations were available, they differed substantially in the details of their semantics, and actual adherence to the emerging language standards was partial.
This provided serious challenges to computer algebra system implementors, as application demands did not allow inefficient emulation or compatibility layers.

In the platform consolidation period,
the set of suitable widely used environments simplified.
Computer architectures were for the most part using 8-bit bytes, memory was byte addressable, and the \sys{Unicode}, {\sc Ieee} 754 and \sys{XML} standards could be relied upon for the most part for
data representation.
The {\sc posix} standard provided common abstractions across operating systems, though bespoke memory management was still widely used.
Most systems or stand-alone programs were implemented in \sys{Lisp}, \sys{C}, \Cpp{} or Fortran, all of which had mature language standards with mostly compliant implementations.
Freely available cross-platform implementations meant implementors could rely on one particular implementation (\textit{e.g.} \sys{g\texttt{++}} or \sys{clang/llvm}).  The Intel instruction architecture was dominant, though that is no longer the case.
Today, older legacy systems
or modern embedded systems may have diverse architectural considerations that can be handled by earlier methods.

We are now in the mathematical consolidation period where
the main remaining portability issues are 
at the levels of mathematical definitions and algorithm encoding.
These come into play when it is desired to share mathematical data or computations across different computer algebra systems and mathematical proof assistants.
The issue of consistency of mathematical definitions has existed since well before the 
digital computer era.  
Seemingly slight differences in the meaning of 
specific mathematical symbols can make what is an identity or theorem in one system be
false in another system.
Examples of issues that arise include sign conventions, normalisation factors, branch cut choices for inverse complex functions, and foundational logics.
The issue of algorithm encoding is that much of mathematical knowledge is captured in the form of
programs written in the application-level programming language of a handful of systems such as \sys{Magma}, \sys{Maple}, \sys{Mathematica}, \sys{Reduce} and \sys{SageMath} (for computer algebra) or \sys{Mizar}, \sys{Coq}, \sys{Isabelle/HOL} or \sys{Lean} (for proof assistants).  Here, not only may the details of definitions differ, but the encoding of algorithms is by-and-large completely incompatible.

We see that while the low-level portability issues have reached a stage of mature solution,
the issues of mathematical data representation and definition specification remains an active topic of research.

We reach a number of conclusions from our study:

The first conclusion is that portability for computer algebra has many dimensions:
portability from platform to platform,
      for the same platform but over time as either platform or needs and opportunities change,
      and for the same computer but porting the intellectual content of an algorithm or library to a new base.

The second conclusion is that
portability is not automatically easy, and the main challenges have altered over time-scales that are relevant to some software.  
As outlined above, we began in the 1960s with a small number of wildly different machines, but software that was simple by today's standards.
Over the next decades, into the 1990s, many new platforms were introduced and standardisation of languages and interfaces was a work in progress.
We have reached a point where there are fewer machines that are relevant again, and much standardisation to
help, but software is much more ambitious. 
There are, however, a variety of portability issues where the originator of the code has not been sloppy, but the world has changed beneath their feet.  
For example,  systems are now concerned with security or protecting intellectual property, resulting in uncontrolled data relocation, unwritable code segments, code signing or license managers.   Compilers with aggressive code optimization may exploit all behaviour allowable by the language standard, invalidating previously reasonable assumptions about 2-s complement arithmetic or instruction ordering.  So things are sometimes tough again.

The third conclusion is that even when there are a great many very different platforms that must be supported, it is possible to do so elegantly and efficiently with a carefully constructed, small portability layer.

The fourth conclusion is that freely available algorithmic code that runs atop a commercial platform may feel locked down if it is complex.
Arranging that it is portable to different worlds (either commercial or open source) is a long way from automatic. 
So the fact that somebody has a paper describing how they can do something you like may or may not help.
And if they then cease to be available to support what they had done the progress they made may be lost. So portability there is critical!
     
The final conclusion is that the people who care about
portability are mostly invisible, but rest of the world depends on the work they do. 
Take, for example,  the many many who keep Linux afloat. For (much) smaller projects there is a (much) smaller pool of expertise, but maintenance and porting are very close neighbours. We can not offer magic bullet solutions to either, but feel that highlighting the issues is important.

\bibliographystyle{splncs04}
\IfFileExists{IfExistsUseBBL.tex}{%

}{%
\bibliography{main.bib}
}
\end{document}